\documentclass{mem}
\usepackage{natbib}\usepackage{txfonts}\usepackage{balance}
\usepackage{graphicx}
\usepackage[breaklinks,dvipdfm]{hyperref}
\idline{83}{1}
\begin{document}
\def\teff{$T\rm_{eff }$}
\def\kms{$\mathrm {km s}^{-1}$}

\title{
  GRBs by thin persistent precessing  lepton Jets: the long life GRB110328 and the Neutrino signal
}

   \subtitle{}

\author{
D.Fargion  }

  \offprints{D.Fargion}

\institute{
Physics Department. and INFN Rome1, Sapienza Univ., Ple A.Moro 2,00185,
Rome, Italy
\email{daniele.fargion@roma1.infn.it}
}

\authorrunning{Fargion}

\titlerunning{GRBs Jets and Neutrinos}

\abstract{Gamma Ray Burst sources are apparently evolving around us in a harder and brighter samples at far and far redshift. The average output may range  from a near Supernova (nearest events) output to a billion time that power for most distant events. Such a tuned evolution around us is not an anti-copernican signature. It is a clear imprint of a off-axis (nearest sources) beaming versus a rarest in-axis blazing (far redshift sources) by a thin relativistic beam (Lorentz factor up $10^4$ or above, micro-nano steradian solid angle). The main consequence is the rarer and rarer  presences of hardest gamma events (hundreds MeV, GeVs, tens GeVs), nearly one over a twenty, observed with difficulty at largest redshift inside their thinner beamed jets. For this reason these rarest tens GeV beamed events, even observed by EGRET and Fermi, are hardly seen at hundred GeV by Cherenkov telescope (Magic,Hess,Veritas) on Earth. For the same reason and because tens GeV neutrino energy is below Icecube thresholds (threshold to hundreds GeV) we  have not been observed yet a neutrino GRB. However if the GRBs primaries contains  tens GeV neutrino traces (at comparable GRB gamma rate) their  presence  may rise in few years at  Deep Core (a more dense array inside ICECUBE) detector whose lower threshold, ranges just one or few tens GeV energy. Moreover the very recent X ray persistent GRB110328 or( J164449.3 transient event), whose   understanding was  first associated to a cannibal star AGN eating \cite{Shao}, but now \cite{Bloom},\cite{Zauderer} , to a cannibal AGN feeding beamed jet, it is,  more naturally consistent to our  GRB (by few solar mass compact source, not an AGN) spinning, precessing and  blazing model jet, whose geometry is more  aligned and stable (than other GRBs)  to us: its decay law and its average output is fully consistent with our earliest proposals.
}
\maketitle{}

\section{Introduction}

  %%%%%%%%%%%%%%%%%%%%%%%%%%%%%%%%%%%%%%%%%%%%%%%
  The gamma ray burst apparent average isotropic power versus their red-shift
of all known GRB has been published recently, see ref.\cite{Fa_Da_2010}. As shown in  Fig.\ref{fig2} this distribution calls for an unrealistic (or tuned) Gamma Ray Burst evolution around us or, better, it just probes the need of a very thin gamma precession-jet model for GRBs, where beaming geometry explain the puzzle.
These precessing and spinning gamma jet are originated by Inverse Compton and-or Synchrotron
Radiation at pulsars or micro-quasars sources, by their parent ultra-relativistic leptons and final electron pairs. These GRB jet sources are compact neutron stars or Black holes each of few solar masses
born at peak activity  during a Supernova spherical explosion. These Jets are
most powerful at Supernova birth, blazing, once on axis, to us and flashing as GRB events, decaying  their jet power in hours-day scale time and surviving as late dimmed SGRs Jets.
The trembling of the thin jet (spinning, precessing, bent by magnetic fields and companion NS,BH or disk), its tiny aperture and solid angle
(a part over a million-billion of a steradian) explains naturally the observed erratic multi-explosive structure of different GRBs,
as well as its rare re-brightening or its precursors.
In our model to make observed GRB-SN in nearly energy equipartition, the jet
must be very collimated $\frac{\Omega}{\Delta\Omega}\simeq
10^{8}$-$10^{10}$ (\cite{FaSa95b, Fa99, DaF05}) explaining why
apparent (but beamed) GRB luminosity $\dot{E}_{GR-jet}\simeq
10^{53}$-$10^{54}$ erg $s^{-1}$ coexist on the same place and
similar epochs with lower (isotropic) SN powers
$\dot{E}_{SN}\simeq 10^{44}-10^{45} erg s^{-1}$. In order to fit
the statistics between GRB-SN rates, the jet must have a decaying
activity ($\dot{L}\simeq (\frac{t}{t_o})^{-\alpha}$, $\alpha
\simeq 1$): it must survive not just for the observed GRB duration
but for a much longer timescale, possibly thousands of time longer
$t_o\simeq 3\cdot 10^4\,s$. The late stages of the GRBs (within the same
decaying power law) would appear as a SGRs: indeed the same law
for GRB output at late time (thousand years) is still valid for
SGRs. The jets are precessing (by binary companion or inner disk asymmetry) and
decaying by a decay law  estimated \cite{Fa99}:
$I_{apparent}$= $ I_{jet}\cdot (\Delta \Omega(t))^{-1} $.
\begin{equation}\label{eq1}
I_{jet}\,=\,I_{1}\;\left( \frac{t}{t_{0}}\right) ^{-\alpha }\simeq
10^{45}\left( \frac{t}{3\cdot 10^{4}\,s}\right) ^{-1}\;\;erg\,s^{-1}
\end{equation}
Where the beaming solid angle is
$(\Delta \Omega(t))^{-1}$= $\theta_1(t)^{-2}$= $[{\theta_{1m}^2 +\theta_{var}^2} ]^{-2}$,
where the minimal average opening angle (main jet-observer)
 $\theta_{1 m}\simeq \frac{1}{\gamma_e}$, $\gamma_e \simeq 10^4$ and the variable jet spinning-precessing angle $\theta_{var}$, rules the whole erratic GRB variability.
%%%%%%%%%%%%%%%%%%%%%%%%%%%%%%%%%%
The huge apparent GRBs luminosity (up to $10^{53}-10^{54}$
erg s$^{-1}$) may be due to highest collimated on-axis blazing jet (trembling and precessing) of the Jet (at SN output power).
 The relic pulsar (or BH) source of the Jet must reflect its spin ($\omega_{psr}$)
frequency in angle $\theta_1(t)$ evolution if his angular momentum axis is
not in general coincident with the gamma jet axis. This fast spinning will, usually, imprint
the ``trembling'' millisecond behavior of most in-axis structured rapid GRBs. Finally the
possible anisotropy of the jet system (related for instance to its own
different inertial momentum, orthogonal and parallel, to the spin axis $%
I_\perp$, $I_{||}$) would modulate by nutation the beam-observer angle $%
\theta_1$ by an angular velocity $\omega_N \sim \omega_{psr}\, \frac{I_\perp
- I_{||}}{I_{||}}$. The combined multi-precessing and spinning beam angle
will describe in the sky a multiple cycloidal (or epicycloidal) trajectory
(almost stochastic) described (in present approximation) by
$
\theta_1(t) = \sqrt{[\theta_{1 m} +\theta_{psr}\cos(\omega_{psr}t )+\theta_N\cos(\omega_{N} t)]^2 +}\\
$
$
\overline{+\;\;[\omega_{b} t + \theta_{psr} \sin (\omega_{psr} t) + \theta_N \sin (\omega_{N} t )]^2} \;\;.
$
%%%%%%%%%%%%%%%%%%%%%%%%%%%%%%%%%%
Additional free parameters to be applied (to fit at best the GRB behavior) in both $\sin (\omega t)$,  $\cos (\omega t)$ terms are their initial phase  pulsar spinning $\phi_{psr}$, and its nutation phase $\phi_{N}$ .
Late GRBs  jets are surviving hundreds year later after the SN-GRB birth as SGRs applying the same eq.\ref{eq1}. GRB blazing occurs inside the observer thin cone of view only a fraction of seconds duration times because of the this spinning jet and its thin beam;  because of  relativistic synchrotron (or IC) laws the jet angle is thinner and thinner for harder and harder gamma spectra but the angle it is wider beamed for soft X band. This explain the longer and longer soft X afterglow and its eventual X precursor appearance. GRB apparent brightening is so well correlated with its hardness (the Amati correlation) because better in axis means more luminosity and harder photons. This explain the wider and longer X GRB afterglow life  with respect to harder and fast gamma GRB structures. The jet peripherals casual blazing explains  the ($6-20\%$) rare presence of (otherwise mysterious) X-ray precursors, events
well before the (apparent) main GRB explosion (for us, a better blazing alignment).  The jet is fed by lepton pair jet (probably first  PeVs muons and later on TeVs and tens GeVs electron secondaries). The absence of the PeV neutrino parent are related to the rarety of the inner harder beaming
 in analogy to GeV gamma rarety in Fermi, EGRET GRBs.
The best detectable GRB neutrino events are therefore the most distant ones, (larger sample, more probable in axis alignment) whose gamma and X-OT afterglow might be even too diluted and red-shifted to be observed.
TeVs neutrino might be too rare to be observed. Tens GeV neutrinos (within a Fermi like power GRB neutrino spectra) might rise in Deep Core in next few years. Because the vertical axis in Deep Core pointing to the North is the less noise area \cite{Fa_Da_2011}, these future correlated GRBs-Neutrino events maybe well recognized.
  %%%%%%%%%%%%%%%%%%%%%%%%%%%%%%%%%%%%%%%
 The well probed Super-novae-GRBs connection since $1998$, often forgotten in fountain fireball models, naturally requires  a thin beaming whose softer external cone
 (as for nearest GRB980425) is explaining the huge diversity between spherical SN output  and apparent coexisting beamed GRB. The presence of a huge population of active jets fit a
wide spectrum of GRB morphology \cite{Giovannelli}.   Last GRB-XRF  080109  extreme vicinity and lowest output maybe understood
  as the external tail of an off-axis GRB jet.  Indeed a similar lesson as earliest GRB980425 or recent GRB060218 tell us that  Supernovas may often contain a Jet senn in different angles. The nearest  jet persistent  activity
    may shine a little off-axis as a soft GRBs or at largest volumes and largest sample  as the brightest and hardest gamma GRB event.
 %%%%%%%%%%%%%%%%%%% \begin{figure}[h]
 %%%%%%%%%%%%%%%%%%%\begin{center}
 %%%%%%%%%%%%%%%%%%%\includegraphics[width=2.5in]{fargion_2011_FINAL_01.Fig00.eps}
 %%%%%%%%%%%%%%%%%%%\caption{The rare NGC 2770 twice SN within a week time: the XRF080109-SN2008D has a statistical deep meaning; these events maybe very common.
 %%%%%%%%%%%%%%%%%%%The long life (days) and soft XRF luminosity, as for the event above, imply a SN-GRB jet whose precessing is observed much off-axis, nearly at widest angle} \label{fig0}
 %%%%%%%%%%%%%%%%%%%\end{center}\label{fig0}
 %%%%%%%%%%%%%%%%%%%\end{figure}
  Because of GRB-XRF  080109 near location (z = 0.0065) it also calls for a
  huge population of such SN-XRF in far Universe, undetected because below the present Swift, Fermi threshold.
    Late stages of these jets may loose the SN traces and appear
  as a short GRB or a long orphan GRB (depending on jet angular velocity and
  view angle, time distance from the SN birth). XRF are just off-axis viewing of the persistent jets of GRBs whose old Supernova are too dim to be observed.
  \section{GRB 110328A: a new model tidal star-AGN cannibalism or just a persistent GRB beaming jet?}
\begin{figure}[h]
\begin{center}
\includegraphics[width=2.4in, height=1.2in]{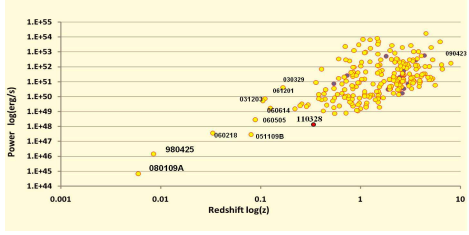}
 \caption{The GRB X-ray luminosity  with most events  updated  up to March 2011.
 The apparent GRBs Luminosity vs red-shift distribution bounded by  a quadratic power, it is mostly due to the quadratic
 distance threshold cut-off and (in higher regions) by the rarer beaming in axis  by largest samples and widest cosmic volumes.
 The spread of magnitude in Luminosity (iso) calls for a thin ($0.001-0.0001$ rad and a micro or nano-sr solid angle) beams. Note the well fit role of persistent GRB110328 understood as a new kind of AGN cannibalism \cite{Shao}. More recent models call for a  jet blazing by AGN, not precessing at all, \cite{Bloom},\cite{Zauderer},\cite{Burrows}. On the contrary a  persistent precessing GRB beamed jet,in stable partial ($\simeq 10^{-2} $ rad, or $0.5^{o}$) collimation to us explain the long-life event. } \label{fig2}
 \end{center}
\end{figure}

  Sw 1644+57/GRB 110328A was discovered by the Swift satellite. Actually it rised earlier (25 March) as a precursor.
It is coincident with an optical source at nearby redshift z = 0.353  as
well as a radio source. The  the luminosity of the flaring X-ray afterglow reaches  $10^{48} $erg s, well explained by a partial beaming of the jet and other GRBs map evolution, see Fig \ref{fig2}; GRB 110328A decayed by nearly 2.5 order of magnitude in nearly $10^{7} s.$, well coexistent  with our (old see eq. \ref{eq1}also in \cite{Fa99}) assumed as inverse linear decay power; the jet beaming increases the apparent luminosity by $3-5$ order of magnitude; for us this variability are due to partial geometry blazing in out axis. The GRB 110328 if it was full in axis it would shine four-six order of magnitude brighter. The time structure is at least tens sec. because the large impact parameter angle;an un-probable better collimation (the source is near , z= 0.3) would show a faster structures, harder spectra brighter flux. The common understanding of this event by many authors  begun as a pure \cite{Shao} star cannibal eating  (almost spherical radiation) to a more AGN-Jet feeding \cite{Bloom},\cite{Burrows},\cite{Zauderer} beamed blazing at fountain angle ($5^{o}-10^{o}$). In analogy of the past GRB Fireball model evolving into  to Fountain Jet Fireball model: from isotropy to a mild beaming. We disagree with both interpretation. These new GRB model (cannibal-AGN jet) are un-probable for different reasons: first the three day precursor followed
  by a silent stage, a very fast variability on 28 March at earliest stages in disagrement with any reasonable early tidal fragmentation process of a star (where one  does expect a slow star disruption and a slow growing feeding of the star debris into  accretion disk and-or AGN jet  regime). The GRB 110328A  shows huge variability are  at very  early time up to 3-4 order of magnitudes in days 2-18 April, unexpected in any
feeding disk and wide angle jet. Secondly the repetitive oscillations are at short minute times, all along, and they are more reminiscent of several long life GRBS structure;
we explain it by a relative large impact angle $\theta_{min}$ shining and blazing to us. Our few solar mass thin, spinning and precessing jet GRB-SGR is quite common, their cannibal jet
is an unique and rarest AGN beamed source, even centered within a kilo-parsec radius from the galaxy core.
The very peculiarity of  of this  GRB, if any, is its stability, its (partial) alignment to us, its detection, within Swift threshold during a long observation life.

  %%%%%%%%%%%%%%%%%%%%%%%%%%%%%%%%%%%%%%%%%%%%%%%%%%

 \begin{figure}[h]
\begin{center}
\includegraphics[width=2.4in,height=1.1in]{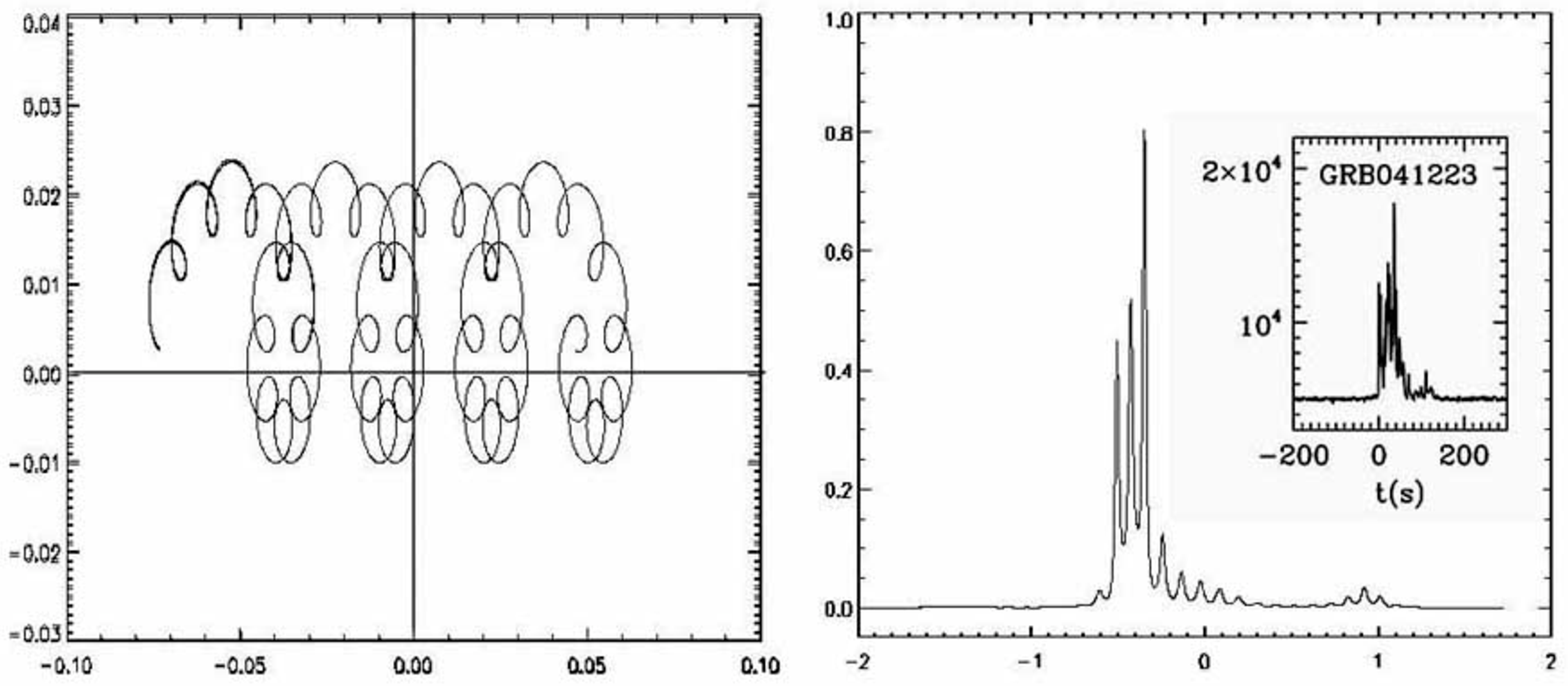}
\caption{The possible simple beam track of a precessing jet to
observer located at origin. On the left, observer stays in (0.00 ;
0.00); the progenitor electron pair jet (leading by IC\cite{FaSa98}
to a gamma jet) has here a Lorentz factor of a thousand and
consequent solid angle at $\sim\mu$ sr. Its consequent blazing light
curve corresponding to such a similar outcome observed in
GRB041223.We assumed nearly thousand Lorentz factor and IC radiation.}\label{fig9}
\end{center}
\end{figure}

\begin{figure}[t]
\begin{center}
\includegraphics[width=2.4in, height=1.1in]{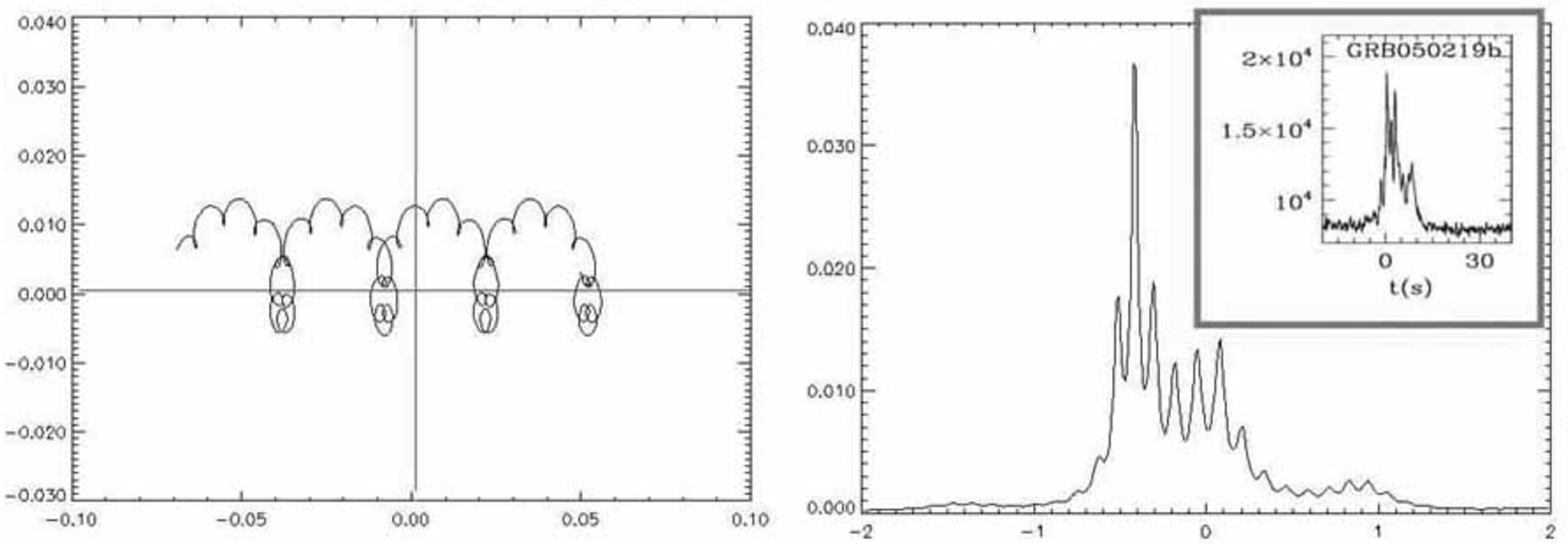}
\caption{Same as in Fig. \ref{fig9}: a precessing jet, as above, its
consequent light curve versus a similar outcome observed in
GRB050219b.} \label{fig10}
\end{center}
\end{figure}

\begin{figure}[t]
\begin{center}
\includegraphics[width=2.5in, height=1.1in]{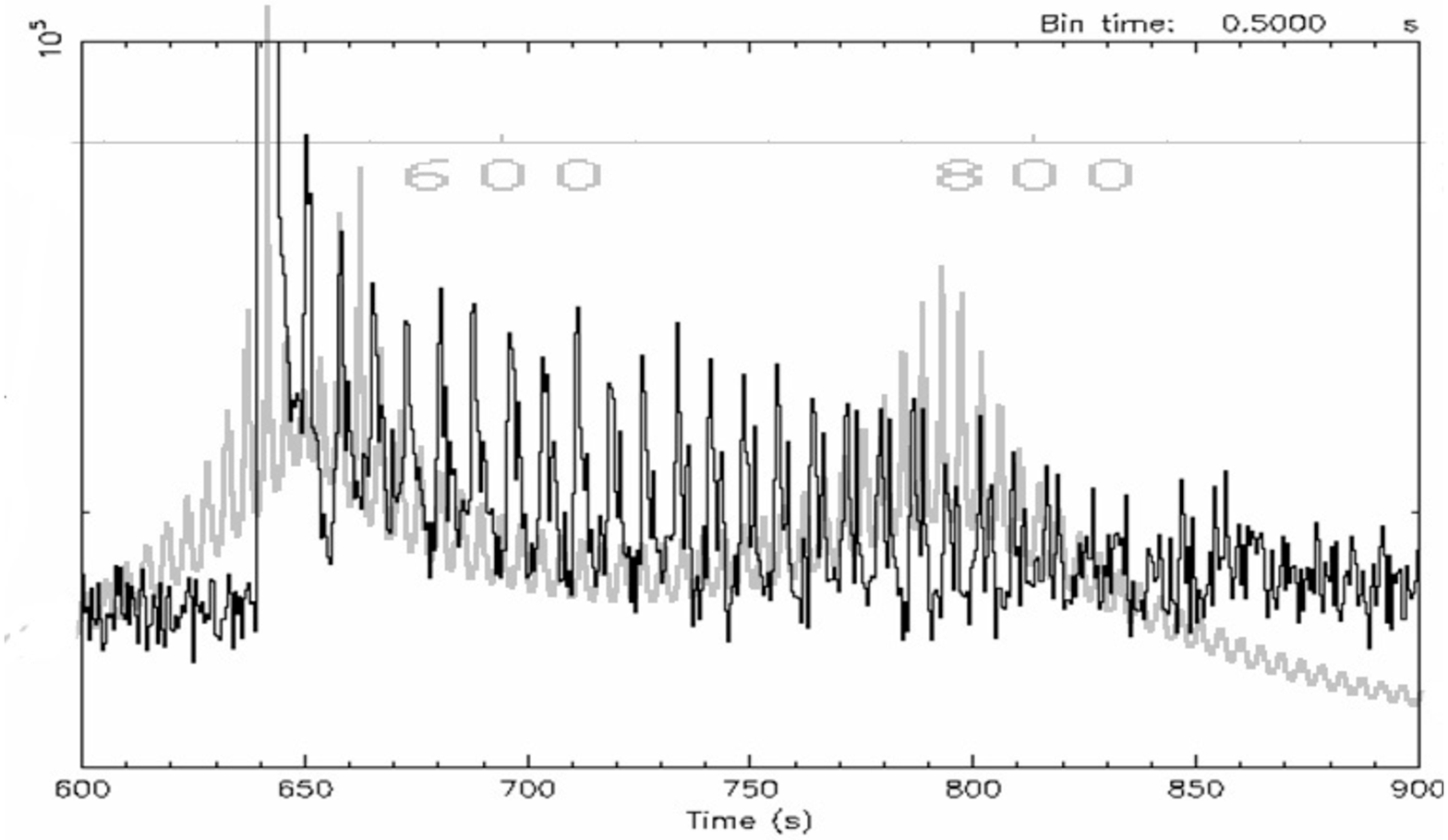}
\caption{Same as in Fig. \ref{fig9}: a precessing jet model by PeV muons, tens or hundred TeV electrons
radiating via synchrotron radiations and the consequent light curve (by spinning and precessing jet) versus a similar outcome observed in huge flare
SGR1806-20.We assumed a Lorentz factor near a billion by PeV electrons} \label{fig11}
\includegraphics[width=2.5in, height=0.7in]{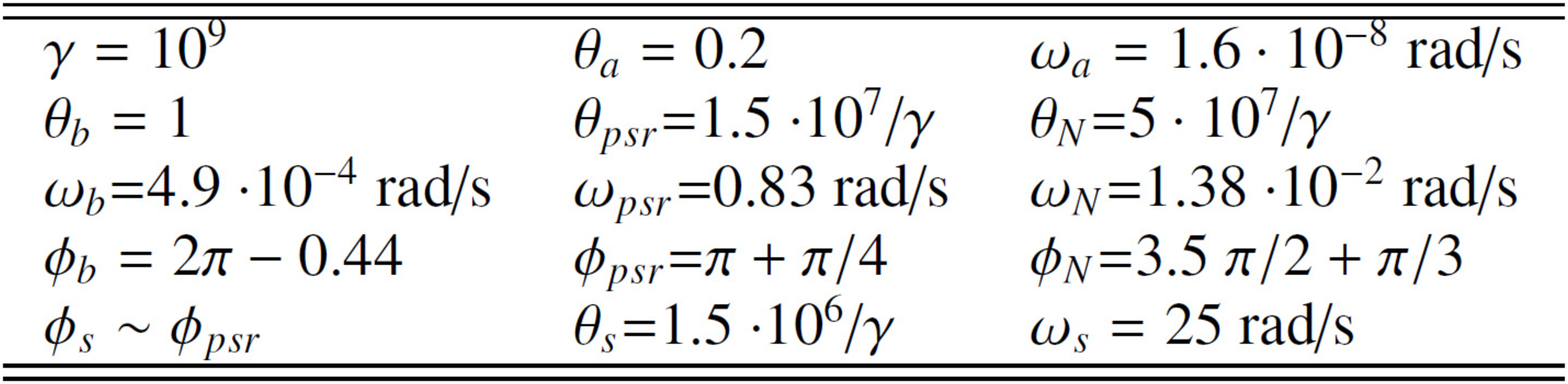}
 \caption{The angular velocity, the spinning and precessing parameters able to fit the rarest and most puzzling SGR1806-20 event described in previous figure \ref{fig11}, occurred by SGR1806-20 on 2004} \label{fig12}
\end{center}
\end{figure}

\begin{figure}[t]
\begin{center}
\includegraphics[width=2.5in, height=1.4in]{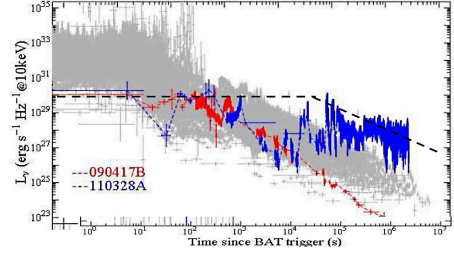}
\caption{The longest live GRB110328 whose X-ray luminosity survived up to day; adapted from \cite{Shao}. The dot line describe an approximate average
 power jet (partially beamed,$\simeq 10^{-2} $ rad, or $0.5^{o}$) collimation to us,  as described by the time evolution in eq. \ref{eq1}.
The event is for us  not a stellar  AGN cannibal eating, \cite{Shao},\cite{Zauderer},\cite{Bloom},\cite{Burrows}, but just (a dimmed) SN-GRB birth in a stable partial ($\simeq 10^{-2} $ rad, or $0.5^{o}$) collimation, blazing to us.} \label{fig1}
\end{center}
\end{figure}
%%%%%%%%%%%%%%%%%%%%%%%%%%%%%%%%%%%%%%%%%%%%%%%%%%%%%%%%%%%5

\section{Blazing jets in GRBs and SGRs}
A key puzzle (or better to say, a lethal question), for one shot popular Magnetar-Fireball
model\cite{DuTh92}, arises for the surprising giant flare from SGR
1806-20 that occurred on 2004 December 27th: if it has been
radiated isotropically (as assumed by the Magnetar
model\cite{DuTh92}), most of - if not all - the magnetic energy
stored in the neutron star NS, should have been consumed at once.
This should have been reflected into sudden angular velocity loss
(and-or its derivative) which was \textit{never observed}. On the
contrary a thin collimated precessing jet $\dot{E}_{SGR-jet}\simeq
10^{36}$-$10^{38}$ erg $s^{-1}$, blazing on-axis, may be the
source of such an apparently (the inverse of the solid beam angle
$\frac{\Omega}{\Delta\Omega}\simeq10^{8}$-$10^{9}$) huge bursts
$\dot{E}_{SGR-Flare}\simeq10^{38}\cdot\frac{\Omega}{\Delta\Omega}\simeq10^{47}$
erg $s^{-1}$ with a moderate  steady jet output power (X-Pulsar,
SS433 like). See simulated gamma SGR1806-20 flare in Fig. \ref{fig12}; This explains the absence of any variation in the
SGR1806-20 period and its time derivative, contrary to any obvious
correlation with the dipole energy loss law.See Ref. \cite{Fargion2006},\cite{DaF06},\cite{DaF05}.
Such a spinning-precessing blazing may also explain the earlier precursor.
The simplest  way to produce the $\gamma$ emission  would be by IC
of GeVs electron pairs onto thermal infra-red photons at Lorentz factor  $\gamma_e \simeq 10^4$. Also
electromagnetic showering of PeV electron pairs by synchrotron
emission in galactic fields, ($e^{\pm}$ from muon decay) may be
the progenitor of the $\gamma$ blazing jet. However, the main
difficulty for a jet of GeV electrons is that their propagation
through the SN radiation field is highly suppressed. UHE muons
($E_{\mu}\geq$ PeV) instead are characterized by a longer
interaction length either with the circum-stellar matter and the
radiation field; thus they have the advantage to avoid the opacity
of the star and escape the dense GRB-SN isotropic radiation field
\cite{DaF05, DaF06}. We proposed \cite{DaF05} that also the emission of SGRs is
due to a primary hadronic jet producing ultra relativistic
$e^{\pm}$ (1 - 10 PeV) from hundreds PeV pions,
$\pi\rightarrow\mu\rightarrow e$, (as well as EeV neutron decay in
flight): primary protons can be accelerated by the large magnetic
field of the NS up to EeV energy. By interacting with the local galactic magnetic
field relativistic pair electrons lose energy via synchrotron
radiation:
$E_{\gamma}^{sync}\simeq4.2\cdot10^6(\frac{E_e}{5\cdot10^{15}\,eV})^2(\frac{B}{2.5\cdot10^{-6}\,G})\,eV$
with a characteristic timescale
$t^{sync}\simeq1.3\cdot10^{10}(\frac{E_{e}}{5\cdot10^{15}\,eV})^{-1}(\frac{B}{2.5\cdot10^{-6}\,G})^{-2}\,s$.
This mechanism would produce a few hundreds keV radiation as it is
observed in the intense $\gamma$-ray flare from SGR 1806-20.
The jet (in PeV synchrotron model) will be not just as the thin cone (of tens GeV IC model), but the jet is a much thinner wider layer of fan-structure. Indeed
the electron pairs are spread by Larmor radius that it is about two orders of magnitude smaller than
the synchrotron interaction length; this  imply that the
aperture of the showering jet is spread in a rare thin fan structure
\cite{Fa97, Fa00-04} by the interstellar magnetic field,
$\frac{R_L}{c}\simeq4.1\cdot10^{8}(\frac{E_{e}}{5\cdot10^{15}\,eV})(\frac{B}{2.5\cdot10^{-6}\,G})^{-1}\,s$.
The solid angle $(\Delta \Omega(t))$ is in this situation, in almost mono-dimensional, of few degree wide and thin:
as thin as the inverse of the Lorentz
factor : $(\Delta \Omega(t))^{-1}$= $\theta_1(t)^{-1}$= $[{\theta_{1m}^2 +\theta_{var}^2} ]^{-1}$,
where the minimal average opening angle (between jet-observer) is: $\theta_{1 m}\simeq \frac{1}{\gamma_e}$, $\gamma_e \simeq 10^{9}$, ($\theta_{1m}\sim 10^{-9} rad)$. In particular a thin
($\Delta\Omega\simeq 10^{-9}$-$10^{-10}$ sr) precessing jet from a
pulsar may naturally explain the negligible variation of the spin
frequency $\nu=1/P$ after the giant flare ($\Delta\nu<10^{-5}$
Hz). Indeed it seems quite unlucky that a huge
($E_{Flare}\simeq5\cdot10^{46}$ erg) explosive event, as the
needed mini-fireball by a magnetar model\cite{DuTh92}, is not
leaving any trace in the rotational energy of the SGR 1806-20, $
E_{rot}=\frac{1}{2}I_{NS}\omega^2\simeq3.6\cdot10^{44}(\frac{P}{7.5\,s})^{-2}(\frac{I_{NS}}{10^{45}g\,cm^2})$
erg. The consequent fraction of energy lost after the flare is
severely bounded by observations: $\frac{\Delta(E_{Rot})}{E_{Flare}}\leq10^{-6}$. More absurd in
Magnetar-explosive model is the evidence of a brief precursor
event (one-second SN output) taking place with no disturbance on
SGR1806-20 \textit{two minutes before} the hugest flare of 2004 Dec. 27th. The thin precessing Jet while being extremely collimated (solid angle $\frac{\Omega}{\Delta\Omega}\simeq10^{8}$-$10^{10}$
(\cite{FaSa95b, Fa99, DaF05, DaF06}) may blaze at different angles
within a wide energy range (inverse of $\frac{\Omega}{\Delta\Omega}\simeq10^{8}$-$10^{10}$). The output
power may exceed $\simeq10^{8}$, explaining the extreme low
observed output in GRB980425 -an off-axis event-, the long late
off-axis gamma tail by  GRB060218\cite{Fargion-GNC}),  respect to
the on-axis and more distant GRB990123 (as well as GRB050904).
\section{ Conclusions}
 \emph{The GRBs are not the most powerful explosions, but just the most  collimated ones.}  Their birth rate
is comparable to the SN ones (a few a second in the observable Universe), but their thin beaming ($10^{-8}$ sr) make them extremely
rare ($10^{-8}$  perfect alignment up to $10^{-4}$, partial beaming), and long life, to observe,  pointing to us at their earliest (days-months after their SN birth) dates.
Sometimes the observation delay makes the SN-GRB connection lost. Sometimes the dust obscuration may  hide the SN. The link with SN
is often guaranteed in long GRB, but the jet connection occurs also for Short GRBs XRF whose explosive supernova is faded away months or years
earlier. In our Universe thousands of GRBs are shining at SN peak power, but they are mostly pointing else
where. Only one a day might be blazing and detect at SWIFT, Agile, Fermi threshold level. Thousand  of billions mini-jets are blazing (unobserved)
as SGRs in the far Universe. The ones in our Galaxy, near enough, maybe revealed  as SGRs. The GRB-SGRs
connection with X-ray-Pulsars make a possible link to anomalous X-Ray pulsar. This  GRB-SGR link to X and gamma pulsar is to be considered as a possible grand unification of the model. The nearest (tens-hundred Mpc) are observable mostly off-axis (because of probability arguments); the last peculiar GRB 110328A does fit the global GRB flux-redshift diagram: see Fig.\ref{fig2}: therefore it is not a too rare AGN beamed star eating, but more probable
 a partial off-axis beaming GRB whose time evolution do confirm our old model. The most distant GRB are seen mostly on
axis (because larger sample and detector threshold ). Therefore the hardest GRB are often at highest redshift.
But the IR cut-off makes this hundred GeVs gamma bounded. Last persistent GRB 110328A just probe the longevity of the jet,because lucky persistence directionality pointing to us.
Hardest GRB  at  most distant redshift, corresponding to hardest hundreds GeV gamma are often obscured by cosmic or inter galactic photon-IR cut off opacity, making hard to observe them at MAGIC-VERITAS-HESS.\emph{For same reasons neutrino GRBs at hundreds GeV up to PeV may be too rare to be observed in ICECUBE while at  few ten GeV neutrino GRB  event may be observed in Deep Core Detector in the  near future. (If GRB  neutrino spectra extends to tens GeV energy in equipartition with the gamma component.)}


\begin{thebibliography}{100}
\bibitem[(Amati,2002)]{Amati}{Amati L et al., A\&A, 390 (2002) 81 }
\bibitem[(Amati,2006)]{Amati2006}{Amati L. et. al.astro-ph/0607148}
\bibitem[(Bloom,2011)]{Bloom}{ Bloom J. S., et al. Science 333, 203 (2011)}
\bibitem[(Burrows,2011)]{Burrows}{Burrows D.et al.Nature:Vol. 476,421,2011}
\bibitem[(Fargion et al,1998)]{FaSa98}{Fargion D.,et al.: Phys.Usp. 41 (1998) 823}
\bibitem[(Lazzati,2006)]{Lazzati}{Lazzati D. et al. astro-ph/0602216}
\bibitem[(Fargion et al, 2010)]{Fa_Da_2010}{Fargion D.,et al.:Mem. S.A.It. Vol. 81, 440}
\bibitem[(Fargion,2011)]{Fa_Da_2011}{Fargion D.,Nucl.Phys.B212:146-153,2011}
\bibitem[(Fargion,1999)]{Fa99}{Fargion D., AA\&SS, 138, 507,(1999)}
\bibitem[(Yonetoku,2004)]{Yo2004}{Yonetoku D.,et al. ApJ, 609, 935, (2004)}
\bibitem[(Stanek,2003)]{Stanek}{Stanek R. et al.Ap. J., 591:L17-L20, (2003)}
\bibitem[(Shao,2011)]{Shao}{Shao L. et al.1104.4685.v1-v3, Ap. J. Letters, 734:L33, 2011}
\bibitem[(Fargion ,2003)]{DaF03}{Fargion D.,Chin.J.Astron.Astr.3,472(2003)}
\bibitem[(Falcone, 2006)]{Falcone}{Falcone A.et al.Astrophys.J.641:1010,(2006)}
%%%%%%%%\bibitem{Woo99}{Woods P.M. et al. 1999, ApJ,527,L47; astro/ph 9909276}
%%%%%%%%\bibitem{Campana}{Campana S. et. al. astro-ph/0603475, A\&A accepted}
%%%%%%%%\bibitem{BB}{Blundell K., Bowler M., Astrophys.J. 622 (2005) L129-L132}
\bibitem[(Fargion et.al,1997)]{Fa97}{Fargion D.et.al.Ap.J. 517,725(1999)}
\bibitem[(Fargion et.al,2002-04)]{Fa00-04}{Fargion D.,Astrophys.J. 570, 909,(2002) \emph{and} Fargion D. et al., Ap.J. 613, 1285, (2004)}
\bibitem[(Fargion et.al,1995)]{FaSa95b}{Fargion D.,Salis A., Astrophysics\& Space Science,231, 191, (1995)}
\bibitem[(Fargion et.al,2005)]{DaF05}{Fargion D.et.al.,Nuovo Cim.28C,809,(2005)}
\bibitem[(Fargion et.al,2006b)]{DaF06}{Fargion D.,et.al,Chin.J.Ast.Ast.6S1,342,(2006)}
\bibitem[(Duncan,1992)]{DuTh92}{Duncan R.,et.al., ApJ, 392, L9,(1992)}
\bibitem[(Fargion,2006)]{Fargion-GNC}{Fargion,D. GNC 4819, 06/02/23}
\bibitem[(Fargion,2006b)]{Fargion2006}{Fargion D.,et al.Proc. Fourth Workshop, Isola d'Elba, Italy,133, World Sc. Publ, (2007) }
%%%%%%%%%%%%%\bibitem{Ghisellini} {Ghisellini G. ,et.al. MNRA.Soc.Lett.375:L36-L40,(2007)}
\bibitem[(Giovannelli et.al.,2003)]{Giovannelli}{Giovannelli F. et.al.Chin.J.Ast.Astr.3,Suppl,1,(2003)}
%%%%%%%%%\bibitem{Med01}{Medina-Tanco, G.A.,   Watson, A.A, 2001, {\em Proceedings of ICRC}, 531.}
%%%%%%%%%\bibitem{Moretti}{Moretti, A.,  2006, $www.merate.mi.astro.it/~moretti/lc_060218.gif$}
\bibitem[(Zauderer et al,2011)]{Zauderer}{Zauderer B.et al.Nature,vol.476,425,(2011)}
\end{thebibliography}
\end{document}